\documentclass[doublecol]{epl2} 

\title{Extended Gutzwiller wavefunction for the Hubbard-Holstein model}

\author{P. Barone\inst{1} \and R. Raimondi\inst{2} \and M. Capone\inst{3,4} \and C. Castellani\inst{4} \and M. Fabrizio\inst{1,5}}
\shortauthor{P. Barone \etal}

\institute{                    
  \inst{1} International School for Advanced Studies (SISSA), 
Via Beirut 2-4,  I-34014 Trieste, Italy\\
  \inst{2} Dipartimento di Fisica "E. Amaldi," Universit\`a di Roma Tre, 
Via della Vasca Navale 84, 00146 Roma, Italy\\
   \inst{3} Istituto dei Sistemi Complessi, CNR, Via dei Taurini 19, 00185 Roma, Italy\\
   \inst{4} SMC, CNR-INFM and Dipartimento di Fisica, Universit\`a "La Sapienza", P.le Aldo Moro
   2, 00185 Roma, Italy\\
   \inst{5} The Abdus Salam International Center for Theoretical Physics (ICTP),
P.O. Box 586, I-34014 Trieste, Italy
}

\abstract{We introduce a new type of Gutzwiller variational wavefunction for correlated 
electrons coupled to phonons, able to treat on equal footing electronic and lattice degrees of freedom.  
We benchmark the wavefunction in the 
infinite-$U$ Hubbard-Holstein model away from half-filling on a Bethe lattice, where we can directly compare with exact 
results by Dynamical Mean-Field Theory. For this model, we 
find that variational results agree perfectly well with the exact ones. In particular the 
wavefunction correctly describes the crossover to a heavy polaron gas upon increasing the 
electron-phonon coupling.}

\begin{document}

\maketitle

The coherent electron motion is strongly slowed down by Coulomb repulsion close to a Mott metal-to-insulator transition. 
In this situation the effects of the electron-lattice coupling may be enhanced, since the lattice has more time to adjust 
following the electron motion. In fact, the Mott transition in real systems is often accompanied by 
substantial structural changes that turn it into a strong first-order one, as it happens in the prototypical 
example of V$_2$O$_3$.~\cite{McWhan} However, despite its relevance in many strongly correlated materials,   
the electron-phonon (e-ph) interaction has been just touched on in the context of the Mott transition.  
This has happened partly because most theoretical efforts have focused mainly on the strong correlation alone, or
on the purely e-ph driven polaron problem,
but also because it is not obvious how to deal with the e-ph coupling once the 
electron-elecron repulsion slackens quasiparticle motion so much that the phonon dynamics becomes comparable or even faster.
For this reason, recent attempts to understand the role of the e-ph coupling 
near a Mott transition have mainly limited to ``exact'' numerical simulations using several techniques, including 
Quantum Monte Carlo~\cite{scalapino,gunnarsson-rosch,misha,kornilovitch}, 
exact diagonalization~\cite{feschke}, Density-Matrix Renormalization Group~\cite{jeckelmann} and Dynamical
Mean-Field Theory (DMFT)~\cite{hewson-coreani,capone2004-sangiovanni2005} or extensions\cite{macridin}. 
However, besides accurate numerical results, it is desirable to have at disposal also 
analytic or semi-analytic approaches, even though approximate, that may provide us with  
simple and intuitive physical insights and may be extended to situations where numerical approaches 
become hardly feasible. 

A popular and simple tool for the study of strongly correlated
electron systems, so far in the absence of e-ph interaction, is the Gutzwiller variational 
approach. In its simplest version~\cite{Gutzwiller}, 
the Gutzwiller approach amounts to variationally modify the relative weights of 
the local electronic configurations of an uncorrelated wavefunction. This variational wavefunction is able to  
describe a Mott transition only in the limit of infinite coordination~\cite{Metzner}, 
where the expectation value of the Hamiltonian can be calculated analytically. In finite-coordination lattices,  
it has been recently argued~\cite{capello} that the Gutzwiller wavefunction may still   
describe a Mott transition once a long-range Jastrow factor is included. 
In this paper we propose an extension of the Gutzwiller wavefunction to account for 
the e-ph coupling, which we label Gutzwiller Phonon Wavefunction (GPW). 
We test the accuracy of this wavefunction in the Hubbard-Holstein model on a Bethe lattice, 
where analytical results can be obtained and compared with ``exact'' DMFT calculations.

We consider an Hubbard-Holstein model
\begin{eqnarray}\label{ham_hh}
H&=&-\frac{t}{\sqrt{z}}\sum_{\langle i,j\rangle,\sigma} c^\dagger_{i\sigma} c^{\phantom{\dagger}}_{j\sigma} 
+ U\sum_i n_{i\uparrow}n_{i\downarrow} +\omega_0\sum_i b^\dagger_i b^{\phantom{\dagger}}_i 
\nonumber \\
&&  + \alpha\omega_0\sum_i (n_i-1)(b_i^\dagger+b^{\phantom{\dagger}}_i)
\end{eqnarray}
where $c^{\phantom{\dagger}}_{i\sigma}\,(c^\dagger_{i\sigma})$ and 
$b^{\phantom{\dagger}}_{i\sigma}\,(b^\dagger_{i\sigma})$ are 
annihilation(creation) operators at site $i$ for
spin-$\sigma$ electrons and for dispersionless phonons of 
frequency $\omega_0$, respectively, $U$ the on-site Hubbard repulsion 
while $\alpha$ parameterizes the e-ph coupling. The hopping $t/\sqrt{z}$ connects nearest neighbor sites on 
a Cayley tree with coordination number $z$. The calculations will actually be performed in the limit $z\to \infty$, 
namely for the so-called Bethe lattice, which has a semi-circular density of states with 
half-bandwidth $D=2t$. The e-ph properties are conveniently parameterized through the adiabaticity parameter 
$\gamma=\omega_0/D$, and the dimensionless coupling  $\lambda=2\alpha^2\omega_0/ D=2\alpha^2 \gamma$.
The choice of a Bethe lattice has the advantage to allow for analytical calculations which can be 
compared directly with DMFT, which is exact in infinite-coordination lattices.
In the DMFT calculations we use exact diagonalization as impurity solver, with  
the conduction bath approximated by a finite number of levels $N_b=9$.    
We refer to Refs.~\cite{capone2004-sangiovanni2005} for more details 
on the practical implementation of this technique to the Hubbard-Holstein model. 
The phase diagram of the Hamiltonian (\ref{ham_hh}) contains several different phases 
depending on the various parameters $\lambda$, $\gamma$, $U/D$ as well as on the electron density. 
In order to simplify the analysis, we will assume a very large, actually infinite, $U/D$, which 
excludes superconductivity. Moreover we will discard the possible occurrence 
of ferromagnetism very close to half-filling.
Under these assumptions, we expect, away from half-filling, an evolution from a strongly correlated metal 
at small $\lambda$ into a gas of heavy polarons at large $\lambda$. Previous DMFT 
calculations~\cite{capone2004-sangiovanni2005} have shown that this evolution is continuous for large (but finite) $U$, 
although the crossover may be quite sharp, while variational Lang-Firsov transformations
 within the slave-boson approach~\cite{perroni2005}  
find instead a first order phase transition between the two metallic phases~\cite{barone}.
We show that polaron formation occurs as a crossover also for $U/D \to \infty$, and that our novel variational approach 
agrees extremely well, also quantitatively with exact DMFT, correctly describing the crossover.

In analogy with purely electronic models~\cite{buneman1998,attaccalite2003},
we introduce a Gutzwiller wavefunction  
that depends also on the phonon coordinates $x_i =(b_i +b_i^\dagger)/\sqrt{2}$ through:
\begin{equation}\label{gutzwiller}
|\Phi \rangle=\Pi_i\, {\cal P}_i (x_i)\,|\Psi_0\rangle,
\end{equation}
where $|\Psi_0\rangle$ is an uncorrelated fermionic wavefunction, here assumed to be 
the non interacting Fermi sea.
The Gutzwiller correlator is defined by 
\begin{equation}\label{correlator}
{\cal P}_i (x_i)=\sum_{\nu=0,1,2}\, \sqrt{\frac{P_{\nu}}{P^{(0)}_{\nu}}}\; 
\phi_{\nu} (x_i)\, |i;\nu\rangle \langle i; \nu|,
\end{equation}
where $|i;\nu\rangle\langle i;\nu| $ is the projection operator at site $i$ onto 
the subspace with electron number $\nu=0,1,2$, 
$P_{\nu}$ are variational parameters while $P^{(0)}_\nu$ are the occupation probabilities
of the Fermi sea, {\it i.e.}, $P^{(0)}_0=(1-n/2)^2$, $P^{(0)}_1=n(1-n/2)$, and 
$P^{(0)}_2=(n/2)^2$, $n$ being the average number of electrons per site.
Finally, $\phi_{\nu} (x_i )$ are normalized 
functions associated to the different local electronic configurations.  
The introduction of this variational freedom is the novelty of our approach. 
We emphasize that the wavefunction (\ref{gutzwiller}) can be easily modified to account for other 
types of e-ph coupling as well as for the phonon dispersion, although that would likely require numerical evaluation 
of the energy and optimization of the variational parameters. 
For instance, phonon dispersion can be included if $|\Psi_0\rangle \rightarrow 
\Phi(x_1,x_2,\dots,x_N)\,|\Psi_0\rangle$, where $\Phi(x_1,x_2,\dots,x_N)$ is a variational many-body 
phonon wavefunction that includes inter-site correlations.

Without loss of variational freedom, we can impose that 
$\int dx_i  \langle\Psi_0 |\, {\cal P}_i^2\, |\Psi_0\rangle =1$ and 
$\int dx_i \langle\Psi_0 |\, n_i\, {\cal P}_i ^2\,|\Psi_0\rangle =n$. 
These two constraints are fulfilled by $P_{0}= d+\frac{\delta}{2}\label{p_0}$, 
$P_1 =  1-2d\label{p_1}$, and 
$P_2 =  d-\delta/2\label{p_2}$, having introduced the parameter $d=(P_0+P_2)/2$ 
and the deviation from half-filling $\delta =1-n$. 
The variational approach amounts to compute the expectation value of the Hamiltonian 
(\ref{ham_hh}) over the state (\ref{gutzwiller}). In infinite-coordination lattices such as our  Bethe lattice the computation becomes  
straightforward thanks to the parameterization in eq.~(\ref{correlator})~\cite{buneman1998,attaccalite2003}, 
and leads to a variational energy per site  
\begin{eqnarray}\label{energy}
E&=&\sum_{\nu=0,1,2}P_{\nu} \langle h_0 (x)\rangle_{\nu} 
+\sqrt{2}\alpha\omega_0 \Big[ P_0\langle x\rangle_0 -P_2\langle x\rangle_2\Big]\nonumber \\
&-&\frac{2|\varepsilon|}{1-\delta^2}|S|^2+U \ P_2.
\end{eqnarray}
Here $h_0 (x)=\omega_0/2\, (-\partial_x^2+x^2)$, 
$\langle\ldots\rangle_{\nu}$ indicates the average over $\phi_{\nu}(x)$,
$|\varepsilon|$ is the Fermi-sea average of the hopping energy per site. The hopping renormalization, $S$, 
is proportional to the overlap between the phonon wavefunctions corresponding to 
two different electronic configurations according to
\begin{equation}\label{hopping_s}
S=\sum_{\nu=0,1} \sqrt{P_{\nu}P_{\nu+1}}\int {\rm d}x \ \phi^*_{\nu}(x)\phi_{\nu+1}(x).
\end{equation}

Minimization with respect to $d$ leads to 
\begin{eqnarray}
&&U+ \Big( \langle h_0(x) \rangle_0+   \langle h_0(x) \rangle_2 -
2 \langle h_0 (x)\rangle_1 \Big)\nonumber\\
&&+\sqrt{2}\alpha\omega_0
\Big(\langle x \rangle_0-  \langle x \rangle_2  \Big)
-\frac{2|\varepsilon|}{1-\delta^2}\frac{\partial |S|^2}{\partial d}\label{mf_d}=0,
\end{eqnarray}
while that with respect  to the phonon wavefunctions yields the following 
non-linear second-order differential equations
\begin{eqnarray}
\frac{\epsilon_0}{P_0} \phi_0& =h_0(x)\phi_0 +\sqrt{2}\alpha\omega_0 x\phi_0
-\frac{2|\varepsilon|}{1-\delta^2}S\sqrt{\frac{P_1}{P_0}}\phi_1  \label{mf_0},&\\
\frac{\epsilon_1}{P_1} \phi_1 &=h_0(x)\phi_1 
-\frac{2|\varepsilon|}{1-\delta^2}\left( S^*\sqrt{\frac{P_0}{P_1}}\phi_0+
S\sqrt{\frac{P_2}{P_1}}\phi_2 \right)\label{mf_1},&\\
\frac{\epsilon_2}{P_2} \phi_2 &=h_0(x)\phi_2 -\sqrt{2}\alpha\omega_0 x\phi_2
-\frac{2|\varepsilon|}{1-\delta^2}S^*\sqrt{\frac{P_1}{P_2}}\phi_1 \label{mf_2},&
\end{eqnarray}
where the $\epsilon_{\nu}$'s are  Lagrange multipliers introduced to enforce the 
normalization conditions of each $\phi_{\nu}(x)$. 
These equations describe a system of forced harmonic oscillators coupled together by a term proportional 
to $\vert\varepsilon\vert$. The latter can be neglected 
in the antiadiabatic regime $\omega_0\gg|\varepsilon|$, so that  
the wavefunctions $\phi_\nu$'s are just displaced 
oscillators whose displacement is associated to the different local electronic configurations.
In the opposite adiabatic regime, $\omega_0\ll|\varepsilon|$, the coupling term is instead dominant. 
This introduces a tight entanglement between the different phonon-wavefunctions and strong
anharmonicity.

As we mentioned we consider here the $U\to \infty$ limit, where  
double occupation is not allowed, hence $P_2=0$ and $P_0=\delta$, $P_1=1-\delta$, leaving us with the two
coupled equations (\ref{mf_0}) and (\ref{mf_1}) for $\phi_0$ and $\phi_1$.
We solve them iteratively by expanding the phonon wavefunctions in eigenstates of $h_0$, $|n\rangle$, 
with energy $n\omega_0$ as $\phi_0 =\sum c_n |n\rangle$ and 
$\phi_1=\sum d_n|n\rangle\label{exp}$. The equations for the coefficients $c_n$ and $d_n$ are
\begin{eqnarray}
\left[\left(\frac{\epsilon_0}{P_0}-n\omega_0 \right)\delta_{nn'}+\sqrt{2}\alpha\omega_0 x_{nn'}
+\frac{2|\epsilon|}{1-\delta^2}P_1d_nd_{n'}^* \right]c_{n'}=0,\nonumber\\
\left[\left(\frac{\epsilon_1}{P_1} -n\omega_0\right)\delta_{nn'}
+\frac{2|\epsilon|}{1-\delta^2}P_0c_nc_{n'}^* \right]d_{n'}=0.\nonumber
\end{eqnarray}
In practice $n \simeq 100$ is sufficient to get accurate values for the ground state
corresponding to the lowest $\epsilon_0$ and $\epsilon_1$.

A key quantity that characterizes the electronic properties is the effective mass $m^*$, 
obtained through $m^*/m = 1-\partial\Sigma(\omega)/\partial\omega$, 
being $\Sigma(\omega)$ the single-particle self-energy which, for infinite coordination, 
is momentum independent. 
Within the Gutzwiller approach, the effective mass is commonly identified as the 
hopping renormalization factor, which, through eq.(\ref{hopping_s}), is given by
\begin{equation}\label{meff}
m^*(\lambda )=m^*(0)\; |\langle\phi_1|\phi_0\rangle|^{-2},
\end{equation}
where the overlap $\langle\phi_1|\phi_0\rangle$  includes the effect of phonons, while 
$m^*(0)$ only the correlation effects. 

\begin{figure}[htb]
\includegraphics[width=8cm]{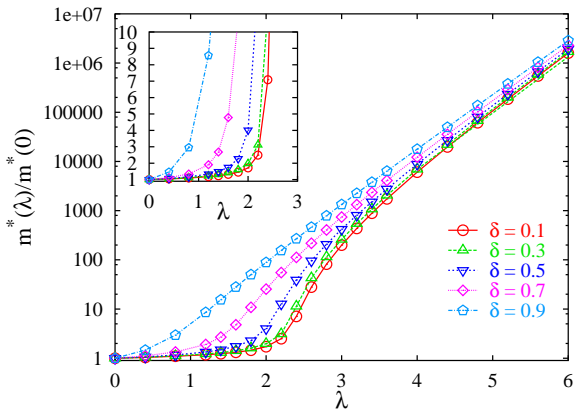}
\caption{(Color online) Effective mass as a function of $\lambda$ and for different 
values of $\delta$ at $\gamma = 0.2$ .
}\label{fig4}
\end{figure}

 \begin{figure}[htb]
\includegraphics[width=8cm]{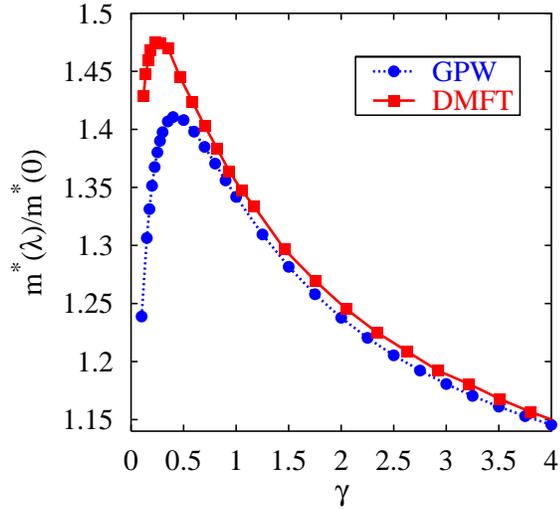}
\caption{(Color online) $m^*(\lambda)/m^*(0)$ as a function of $\gamma$ at $\lambda =1.5$ 
and $\delta =0.3$ in DMFT (squares) and GPW (circles).}
\label{fig5}
\end{figure}

In fig. \ref{fig4} we draw $m^*(\lambda)/m^*(0)$, eq.~(\ref{meff}), for different dopings as 
function of $\lambda$ for $\gamma = 0.2$, in the adiabatic range. The logarithmic scale 
on the vertical axis is used to show how the method reliably works for arbitrary value
of the coupling.
The linear scale in the inset emphasizes the onset of polaronic behavior, identified by a rapid growth 
of $m^*$. As we anticipated, in our calculation the formation of polarons occurs via a crossover,      
in agreement with DMFT calculations.
Approaching half-filling the crossover becomes sharper and it moves to larger $\lambda$'s due to the increase of correlation effects.

To assess quantitatively the accuracy of the GPW, in fig.~\ref{fig5} the 
evolution of $m^*$ as a function of $\gamma$ for $\lambda = 1.5$ and $\delta=0.3$ is 
compared with DMFT. 
Like in DMFT, the variational $m^*$ first increases with $\gamma$ in
the adiabatic regime, reaches a maximum and finally decreases in the antiadiabatic regime. 
The GPW results closely follow DMFT for large $\gamma$'s, and deviate more appreciably   
only in the adiabatic region, although the difference with the exact results is never larger than 20\%. 
Yet, the qualitative behavior is correct for any $\gamma$;  
in particular, as previously mentioned, the polaron-formation remains a crossover also deep inside 
the adiabatic regime.  

\begin{figure}[ht]
\includegraphics[width=8cm]{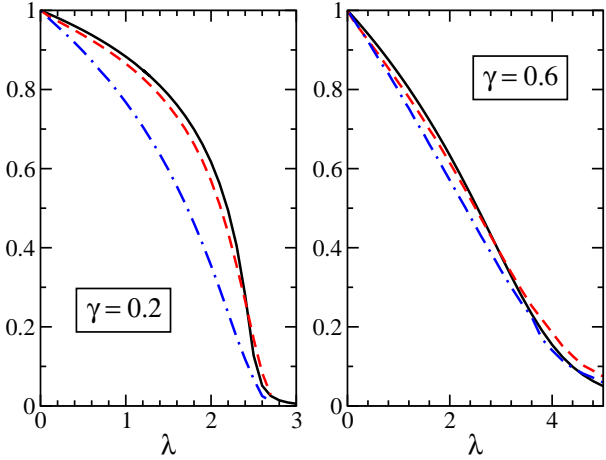}
\caption{(Color online) 
$\vert\langle\phi_0 \vert\phi_1\rangle\vert^2$ 
as a function of $\lambda$ at $\delta = 0.3$ for $\gamma=0.2$ and $\gamma=0.6$ 
in GPW (solid black line) and DMFT (dashed red line), compared with $m^*(0)/m^*(\lambda)$ from DMFT 
(dot-dashed blue line)}\label{fig1}
\end{figure}

The effect of the e-ph coupling on the mass renormalization is hidden in the variational calculation 
into the overlap $\langle\phi_1\vert\phi_0\rangle$. 
In fig.~\ref{fig1} we plot $\vert\langle\phi_1\vert\phi_0\rangle\vert^2$ as a function of $\lambda$  
for two values of $\gamma=0.2,\,0.6$ within the adiabatic region, where the quantitative agreement
with DMFT is poorer. Here we compare with the same quantity from DMFT, and
with $m^*(0)/m^*(\lambda)$ calculated in DMFT from the derivative of the self-energy. Obviously within DMFT eq. (\ref{meff}) 
has no reason to hold.
In both cases the variational wavefunction is able to reproduce extremely well the 
exact value of the overlap. What is clearly different in the two cases is the ability of GPW to
reproduce the results for the effective mass. While for $\gamma=0.6$ the agreement is very good
also as far as $m^*(0)/m^*(\lambda)$ is concerned, which also means that eq. (\ref{meff}) is almost valid within
DMFT, for the smaller value of $\gamma$ $m^*(0)/m^*(\lambda)$ and $\vert\langle\phi_1\vert\phi_0\rangle\vert^2$
are quite different, and the accurate value of the overlap is not translated into an equally
accurate estimate of the effective mass.
This result is consistent with the spirit of our variational approach which is not in principle
accurate for a quantity related to excitations, such as $m^*$, while it correctly
reproduces static quantities such as the overlap. Only when the phonon contribution to the mass
renormalization $m^*(\lambda)/m^*(0)$ can be expressed in terms of the
overlap, the method becomes very accurate also for this quantity.

\begin{figure}[ht]
\includegraphics[width=8cm]{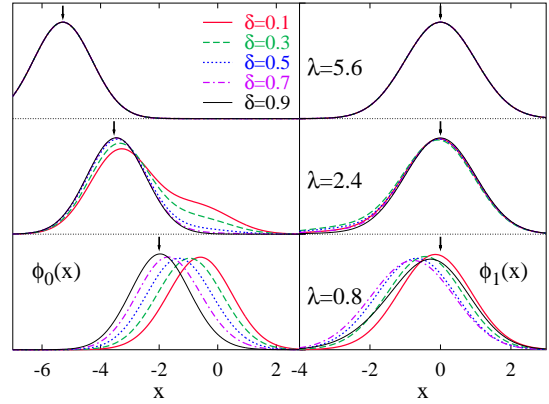}
\caption{(Color online) Evolution of wavefunctions as a function 
of doping for different values of $\lambda$ at $\gamma =0.2$. 
The arrows indicate the displacements in the atomic limit.}\label{fig2}
\end{figure}

Finally, let us discuss in some details the evolution of the phonon wavefunctions, which are  
easily accessible within the variational approach. 
In fig. \ref{fig2} the behavior of $\phi_0(x)$ and $\phi_1(x)$ is shown as a function of doping 
$\delta$ at $\gamma = 0.2$ and for three different values of $\lambda$, 
representative of the weak to intermediate to strong-coupling crossover. At $\lambda =5.6$,
deep into the polaronic regime, the phonon wavefunctions are basically harmonic oscillators 
with a displacement very close to the atomic limit values  
(marked by the vertical arrows in the figure), and with a negligible doping dependence.
Similarly, when the e-ph coupling is relatively small ($\lambda =0.8$),  
the wavefunctions are still similar to displaced  harmonic oscillators, although the 
doping dependence is more pronounced. 
The most interesting behavior occurs for $\lambda =2.4$, which lies in the crossover 
region of fig. \ref{fig1}. Here the shape of the phonon wavefunction 
strongly deviates from the harmonic oscillator and the doping dependence is strong.
In particular, $\phi_0(x)$ develops a shoulder 
besides the main peak. The position of the shoulder roughly corresponds to the maximum of $\phi_1(x)$. 
It is exactly this deformation that makes the continuous evolution as a function of $\lambda$ possible.
We notice that a similar behavior for the evolution of the phonon wavefunction is obtained by means of an exact numerical solution of
the polaron problem on a two-site lattice  in the Holstein model\cite{ranninger92}. 

In this paper we have introduced a generalization of the Gutzwiller variational approach to account 
for the e-ph coupling besides strong electronic correlations. 
We have benchmarked the variational wavefunction in the infinite-$U$ limit of the Hubbard-Holstein model 
on the Bethe lattice, where the variational problem can be solved analytically and directly 
compared with exact DMFT calculations. We have found that the variational results are in perfect 
qualitative, and to a large extent also quantitative, agreement with the exact ones. 
We emphasize that the variational wavefunction (\ref{gutzwiller}) treats on equal footing 
both electronic and phononic coordinates, without assuming either a Born-Oppenheimer scheme 
or an anti-adiabatic one. From this viewpoint, it may provide interesting insights about the 
way retardation effects due to the lattice dynamics show up in the ground state wavefunction, 
especially for strongly correlated systems where the Gutzwiller approach 
is expected to work better.

\acknowledgments
We acknowledge precious discussions with S. Ciuchi and G. Sangiovanni.
This work has been supported by MIUR Cofin 2004 and 2005 and by CNR-INFM.

\end{document}